\documentclass[aps,prl,twocolumn,groupedaddress]{revtex4}

\usepackage{graphicx}

\begin{document}


\title{Spin-coupled  double-quantum-dot behavior inside a single-molecule transistor}



\author{A. Bernand-Mantel}
\author{J. S. Seldenthuis}
\author{A. Beukman}
\author{H. S. J. van der Zant}
 \email{h.s.j.vanderzant@tudelft.nl}
\affiliation{Kavli Institute of Nanoscience, Delft University of Technology, P.O. Box 5046, 2600 GA, The Netherlands}
\author{V. Meded}
  \author{R. Chandrasekhar}
   \author{K. Fink}
     \author{M. Ruben}
      \email{mario.Ruben@kit.edu}
       \author{F. Evers$^{1}$}
        \email{ferdinand.evers@kit.edu}
 \affiliation{Institute of Nanotechnology, Karlsruhe Institute of Technology, D-76021 Karlsruhe, Germany}
 \affiliation{$^{1}$Inst. f. Theorie der Kondensierten
  Materie, Karlsruhe Institute of Technology, D-76128 Karlsruhe, Germany}


\date{\today}

\begin{abstract}
We report on the observation of Kondo and split Kondo peaks in
single-molecule transistors containing a single spin transition molecule with
a Fe$^{2+}$ ion. Coulomb blockade characteristics reveal a
double quantum dot  behavior in a parallel configuration, making our system a
molecular equivalent to a semiconducting double-quantum-dot system. As the gate voltage is increased the charging of the second dot by an additional electron induces a splitting of the Kondo peak. We discuss possible origins of this effect including a spin transition into a high-spin state.
\end{abstract}

\pacs{73.23.-b, 73.63.Kv, 85.65.+h}

\maketitle

The Kondo effect, originally observed in diluted magnetic alloys is induced by the interaction between a local spin and conduction electrons \cite{Kondo64}. In a quantum dot(QD), this effect gives rise to a zero-bias conductance peak (ZBP) for charge states with a total spin S=1/2 \cite{Goldhaber98,Cronenwett98}. When a magnetic field is applied, the degeneracy of the spin state is lifted by the Zeeman effect and a splitting of the ZBP is observed \cite{Costi00}. In the absence of a magnetic filed, a split ZBP corresponding to a bias-induced transition from singlet to triplet state can also be observed for charge states with an integer spin \cite{Sasaki00,Nygard00,Osorio07}. Another system where the Kondo effect has been investigated is the two impurities system originally described by Jones et al. \cite{Jones88}. For example, split ZBPs have been observed in double QD systems where two dots interact via a conducting region or via direct tunneling \cite{Jeong01,Craig04}. The exchange interaction between the spin of the two dots lifts the degeneracy of the ground state. Besides semiconducting double QDs, this effect has been observed in only a few other systems: in  metallic dots containing magnetic impurities \cite{Heersche06} and in scanning tunnelling microscopy experiments with two interacting  magnetic atoms \cite{Otte09}. In addition to the fundamental interest in studying the Kondo effect on single impurities, some of those works have been motivated by the general goal of addressing and manipulating single spins with possible application in quantum information processing \cite{Loss98}. Particularly promising systems towards this direction are magnetic molecules in which the exchange coupling between the spins can be manipulated by charging the molecule with an electric field \cite{Lehmann07}.

In this letter, the Kondo effect is used to characterize the exchange interaction inside a single-molecule transistor (SMT) containing a spin transition (ST) molecule. This type of compound supports a transition between a low-spin (LS) and a high-spin (HS) state, which can be induced by a change in temperature, pressure, magnetic field or by light irradiation \cite{Gutlich04}. We have studied a ST molecule containing a Fe$^{2+}$ metal ion connected to two ligands : [Fe-(L)$_{2}$]$^{2+}$[L=4'-(4'''-pyridyl)-1,2':6'1''-bis-(pyrazolyl)pyridine] (referred to as [Fe-(L)$_{2}$]$^{2+}$) (see Fig. 1(a)). This molecule shows a reversible thermally driven spin transition at 286 K in its bulk crystalline form \cite{Rajadura06}. This type of ST molecule is particularly suited for studies in a SMT geometry since the charge state of the molecule can be modified by applying a potential on the gate electrode. According to recent ab-initio calculations \cite{Ferdinand}, this charging may drive the  ST from LS to HS in analogy to charge transfer ST compounds \cite{Adams93}. 
\begin{figure}[h!]
\includegraphics[width=9cm]{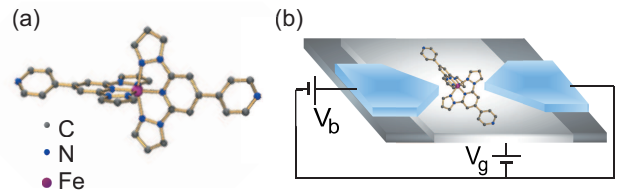}
\caption{\label{schem}Schematic representation of: (a) the molecule [Fe-(L)$_{2}$]$^{2+}$[L=4'-(4'''-pyridyl)-1,2':6'1''-bis-(pyrazolyl)pyridine] and (b) the the transistor geometry.}
\end{figure}

To connect the molecule, we fabricate nanometer sized gaps by a controlled electromigration process on a thin Au wire deposited on top of an oxidized Al gate \cite{Osorio08}. Electromigration and the subsequent self breaking of the wire is carried out at room temperature in a solution containing the molecule; the solvent is subsequently evaporated. If a molecule is trapped in the gap a three terminal device is formed in which the molecule connects via tunnel barriers to source and drain electrodes  and couples  electrostatically to an external gate (see Fig. 1(b)). 

We present in Figs. 2(a) and 3(a) the differential conductance over  bias and gate voltage obtained at $T=1.5$~K for two samples, A and B. The data exhibits  classical behavior of single-electron transport through a nano-object with a single relevant energy level. The extracted addition energies E$_{add}>$100 meV and gate couplings $\beta=C_{g}/C_{tot}\approx0.02$ ($C_{g}$: gate-dot capacitance; $C_{tot}$: total capacitance of the dot) are typical of transport through a single molecule \cite{Osorio08}. The tunnel couplings  for samples A and B can be estimated from the Lorentzian broadening  $\Gamma$ of the Coulomb peaks :  $\Gamma_{A}\approx$60 meV and  $\Gamma_{B}\approx$10 meV. A ZBP, characteristic of the Kondo effect, is observed for both samples in a pronounced regime of gate voltages. For sample A, the Kondo temperature  extracted from the temperature dependence of the zero-bias conductance of the Kondo peak \cite{Costi94}  presented in Fig. 5 is $T_{KA}\approx$ 70 K. For sample B, a temperature T$_{KB}\approx$ 25 K is deduced from its full width at half height. 
\begin{figure}[h!]
\includegraphics[width=8cm]{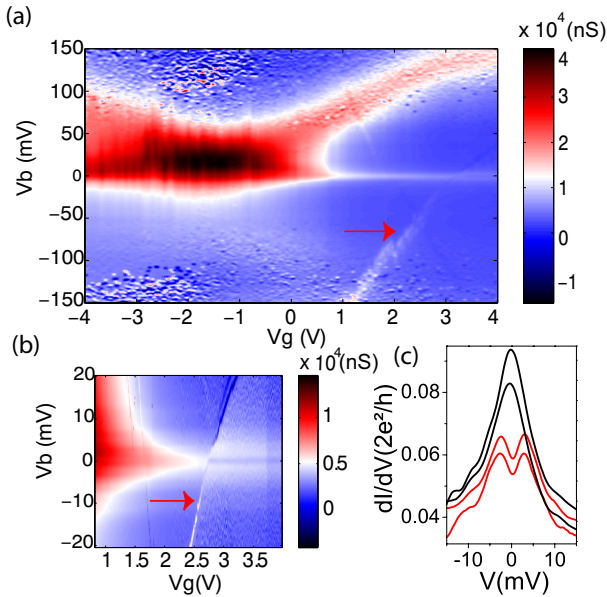}
\caption{\label{SA}  (a)  Differential conductance of sample A versus source-drain ($V_b$) and gate ($V_g$) voltages. (b) Zoom of the differential conductance plot. (c) Differential conductance traces (from top to bottom) versus $V_b$ taken at gates voltages of respectively 2, 2.4, 2.9 and  3.3V. The black (red) traces correspond to gates voltages on the right (left) hand side of the diagonal line. }
\end{figure}
\begin{figure}[h!]
\includegraphics[width=7.5cm]{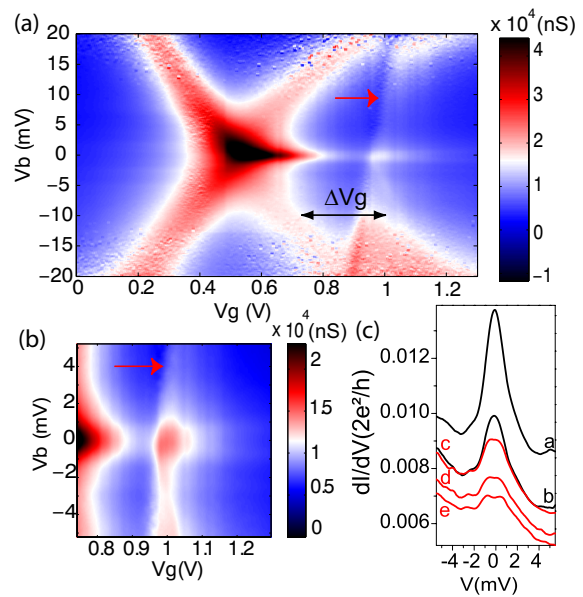}
\caption{\label{SB} (a)  Differential conductance of sample A versus source-drain ($V_b$) and gate ($V_g$) voltages. (b) Zoom of the differential conductance plot (c) Differential conductance traces (from a to e) versus $V_b$ taken at gate voltages of respectively 0.8, 0.9, 1.15, 1.2 and  1.25 V. The black (red) traces corresponds to gates voltages on the right (left) hand side of the diagonal line. }
\end{figure}

In addition to this typical SMT behavior we observe a particular feature visible as a diagonal line (indicated by a red arrow) in the differential conductance plots of Figs. 2(a,b) and 3(a,b). In Fig. 3(a) on the right hand side of this line, a shift $\Delta V_{g}$ of the Coulomb diamond edges in gate voltage is present, which finds a natural interpretation as a response to a modification of the electrostatic environment. This feature can be reproduced by a Coulomb blockade model which is based on a scenario with two parallel dots: current flows through one QD, the "transfer dot'"  in the presence of a second dot, the "spectator dot"(see Fig. 4(a)).   This second dot is electrostatically coupled to the first dot and as its tunnel coupling to one of the electrodes is very weak; there is no net current through it. However when the gate voltage is increased this dot can be charged by an additional electron, thereby, modifying the electrostatic environment of the other dot via their mutual capacitance $C_{M}$. This charging induces an horizontal shift of the  Coulomb diamond edges by $\Delta V_{g}$. Fig. 4(b) shows the result of a rate equation calculation of two single-level QDs capacitively-coupled (see supplementary information for more details), which mimics the situation for sample B. We will discuss later the most likely  physical origin of this double dot scenario. \\
We first focus on the low-bias regime.  A zoom of this region is shown in Fig. 2(b) and 3(b). For sample A the ZBP is clearly split on the right hand side of the diagonal line and two peaks appear at $\pm eV_{b}=\pm \Delta_{A}=\pm 2.75$ meV (see Fig. 2(c)). The temperature dependence of the ZBP and split ZBP are presented in Fig. 5(a,b).  In case of the split ZBP, the zero-bias conductance $G(0)$ exhibits a nonmonotonic temperature dependence (Fig. 5(c)). For sample B the splitting is less pronounced and is visible as a reduction of the zero bias conductance (Fig. 3(c)). When the gate voltage is further increased the splitting becomes more pronounced due to the decrease of the Kondo temperature as we go away from the degeneracy point \cite{Goldhaber981}. For the trace indicated as "e" in Fig. 3(c) we extract a splitting $\Delta_{B}\approx$0.5~meV for sample B.
 \begin{figure}[h!]
\includegraphics[width=8.5cm]{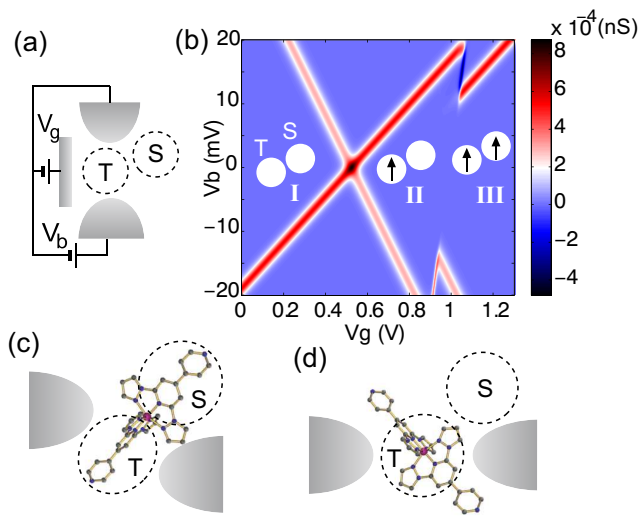}
\caption{\label{SIM} (a) Schematic representation of the double dot system.  (b) Simulation of the differential conductance versus source-drain ($V_{b}$) and gate ($V_{g}$) voltages calculated  for $T=4.2$~K.  (c) and (d) Schematic representation of  molecule-electrode contacts realizing respectivly the {\it intrinsic} ($R_{i}$) and {\it extrinric} ($R_{e}$) double-dot scenario.}
\end{figure}
Split ZBP as well as nonmonotonic temperature dependence of $G(0)$ are reminiscent of spin-spin interactions competing  with Kondo screening, as observed in coupled QDs systems \cite{Jeong01,Craig04} and described theoretically in Refs. \cite{Vavilov05,Simon05}.
With this in mind one might expect that the split ZBP is induced by interactions between the two spins of the double dot also in our case. This explanation would be consistent with the observed transport characteristics in Fig. 2(a) and 3(a), if we assume the occupation of the dots to be as represented schematically in Fig. 4(b). On the left hand side (region I), the transfer dot (1) has an even occupancy (as confirmed by the absence of Kondo peak in this region in Fig. 3(a)).
 When the gate voltage is increased (region II), the transfer dot is
 charged by an additional electron, leading to an odd occupancy,
 S$_{t}$=1/2 and the observation of a Kondo peak (Fig. 2(a) and
 3(a)). In region III the spectator dot is in turn charged leading to
 an odd occupancy S$_{t}$=S$_{s}$=1/2 of both dots, if an even
 occupancy is assumed for the spectator dot on the left hand side of
 the diagram. A sizable spin interaction between the transfer and the
 spectator dot could then explain the experimentally observed peak
 splitting. However, as we will argue below the estimated exchange
 interaction between the two dots is typically too small in order to
 account for the experimentally observed value $\Delta_{A,B}=$0.5-3
 meV. Here the spin transition will come in. 
 \begin{figure}[h!]
\includegraphics[width=8cm]{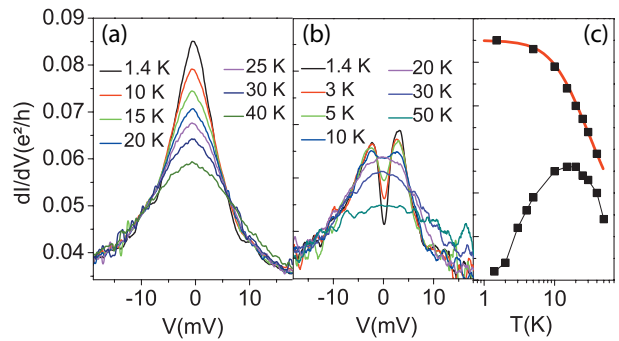}
\caption{\label{TEMP} (a) and (b) Temperature dependence of differential conductance traces of sample A versus V$_{b}$ taken respectively for $V_{g}$=2.2V and $V_{g}$=3.2V. (c) Temperature dependence of the zero bias conductance in (a) and (b) . The red line represents the fit obtained from  the functional form  $G(T)=G(0)[1+(2^{1/s}-1)(T/T_{K})^{2}]^{s}$  \cite{Costi94,Goldhaber981} using $s=0.22$ and $T_{KB}$=70 K.}
\end{figure}

Two possible physical realizations of the double-dot scenario, R$_{i}$
and R$_{e}$, are conceivable (see Fig. 4(c) and 4(d)). In the first
realization, R$_{i}$ , we make explicit reference to the internal
molecular structure of our [Fe-(L)$_{2}$]$^{2+}$ molecule. To support
this scenario, we have carried out ab-inito density functional theory
(DFT) calculations \cite{Ferdinand} (see also supplementary
information) on the molecule. The two bi-pyrazolyl-pyridine ligands
support $\pi$-systems, which are oriented perpendicularly to each
other by the octahedral coordination around the Fe$^{2+}$ metal ion.
 Therefore, left and right ligand states hybridize only weakly
with each other.  In particular,  for the situation of double
reduction (i.e., when two excess electrons are added) our calculations
confirm that each electron  is localized on one ligand only, with very
little excess charge penetrating  the region connecting the two
ligands near the Fe$^{2+}$ ion.  In the asymmetric situation where one
ligand is connected with two electrodes  via  $\pi$-ligand metal-electrode half-sandwich type binding interactions \cite{Orga}, and the other one only to a single electrode (as represented in Fig. 4(c)),  the double dot system
sketched  in Fig. 4(a) is indeed realized. In the alternative way to
produce a double dot system, R$_{e}$,  one could imagine an {\it
  extrinsic} spectator dot, that does not form a chemical bond with
the original molecule, e.g. a second molecule or a small metallic
grain (see Fig. 4(d)).

 We start the discussion with the scenario R$_{i}$ as illustrated in Fig.4~(c). A first estimate for the
 ligand-ligand exchange coupling (with the molecule being in the LS
 state) may be obtained using J$^{\rm L-L}_{i}\sim \varrho^{2}$U
 ($\varrho$=t/E$_{{\rm HOMO}^*}$: relative energy splitting 
 between symmetric/anti-symmetric combinations of ligand
 orbitals; U: effective interaction energy in the near ion region
 of wavefunction overlap, roughly approximated by the
 e$_{g}$-t$_{2g}$-splitting: U$\lesssim \Delta_{\rm oct}$).
 We have typically  $\varrho \approx$0.1-1\%,
 U$\approx$5 eV,  so J$_{i}\lesssim$0.01-1meV. This rough
 estimate is fully supported by our
 {\it ab initio} calculations \cite{Ferdinand} (CASSCF level, see supplementary
 information) that predict for the LS state a weak ligand-ligand
 ferromagnetic exchange coupling J$^{\rm L-L}_{i,{\rm LS}}\approx$0.095 meV.
Essentially, this low value
 is due to the suppression of the tunneling matrix
 elements, $t$, between the two ligand systems. It is only weakly
 modified if a small variation of the 90$^\circ$ angle between the ligands is
 taken into account (0.15meV change at 20$^{\circ}$). To further
 support the theoretical arguments, we can compare these values with observed exchange splittings for LS transition metal
 complexes \cite{Stroh01}, which are in the $\mu$eV range. In conclusion, according to our
 calculations, the scenario R$_{i}$ when assuming a LS state leads to an
 exchange coupling too small to explain the observed splitting
 $\Delta_{A,B}$.
  
  Now, in the second case involving an extrinsic dot, R$_{e}$, the
  direct tunneling between the spectator and the transfer dot
  \cite{Jeong01,Chen04} should be expected to be even smaller than in
  the scenario R$_{i}$ since they are not connected via a chemical
  bond. Hence, also the associated exchange coupling will be
  very small. Of course, in principle there is an additional indirect
  interaction due to the  RKKY coupling via the leads
  \cite{Craig04,Heersche06}.  An accurate estimation of this splitting
  is difficult to obtain as it depends strongly on the spatial
  configuration and geometry of the dots and electrodes.
  However, we argue that the associated exchange
  coupling should anyways be very
  weak. This is because coherent tunneling processes
  connecting the two dots have to take electrons across two barriers: from the
  spectator dot into the leads and then from the leads into the transfer dot (and the same way back to establish coherence). The parametrical estimate ("golden rule" type) for the
  associated coupling is
  J$_{e}^{\rm RKKY}\approx\Gamma_{t}\Gamma_{s}/\Delta_{ts}/(r_{ts}k_{F})^{2} $
  ($\Gamma_{t,s}$:  electrode induced broadening of the molecular
  levels of the transfer and spectator dot; $\Delta_{st}$: mismatch
  of the two coupled energy levels;  $r_{ts}k_{F}$: spatial distance
  of the two contact points of the two dots with the electrodes
  measured in units of the Fermi-wavelength). Assuming
  $\Gamma_{t,s}\approx$10 meV,  $\Delta_{ts}\approx100$~meV and
  $r_{ts}k_{F}\approx 3-10$ typical values of
  J$_{e}^{\rm RKKY}$ lie between 0.01-0.1~meV. Hence we conclude: an
  explanation of the observed split ZBP, which would be based on a
  conventional spin-spin interaction inside a double-dot
  seems hardly consistent with the values of the observed splittings
  $\Delta_{A,B}$.
  
 So far, our analysis was based on the assumption that the molecule
is in the LS state. However, a detailed theoretical analysis suggests
that the LS state may give way against a HS state
when the [Fe-(L)$_{2}$]$^{2+}$ molecule is
doubly charged \cite{Ferdinand}. Accordingly, the molecule could
be in its LS state in region I and II,
while it would be in the HS state in region III.
In HS the 
Fe$^{2+}$ core possesses a spin $S=2$ which is antiferromagnetically
coupled to the spins on the ligands. The ground state
is thus a spin triplet\cite{Ferdinand} 
\footnote{Our CASSCF result indicates that the splitting
 between ground state triplet and the consecutive quintet is at least 53.2 meV corresponding to  
a metal ion-ligand exchange interaction
J$_{i}^{\rm Fe-L}\approx$-13.3meV, in the same range as the one experimentaly observed in similar transition metal complex
\cite{Osorio10,Stroh01}. 
}  
Due to the spin-orbit interaction the triplet splits into a
Kramer's doublet and a singlet state.
In the optimized high-spin geometry of our molecule,
we find that this {\em zero field splitting}, $\Delta E$, is about $0.16$~meV. 
This value will be larger if distortions due to incorporation in the planar device geometry are present
and our preliminary calculations suggests that it can reach
values of several meV. We thus expect that among all the
possible scenarios, R$_{i}^{\rm HS}$ gives rise
to the strongest splittings which approach the experimental values of a few meV.

In summary we have observed a double-quantum-dot behavior in
combination with a split Kondo peak. We suggest that the split ZBP
indicates a charged-induced spin-transition inside the
molecule. To reach a definitive conclusion in favor of this high-spin
scenario, further experiments, which could involve,
e.g., a detailed measurement of the Kondo splitting in magnetic
fields, would be needed. The observation of such a spin-transition
would be an important step towards the development of molecular
spintronics, as it demonstrates the possibility of addressing and
controlling with an electric field individual spin states engineered
at the molecular scale.
\begin{acknowledgments}
Work in part financed by the Dutch funding agency FOM, by the DFG
priority program SPP1243 and the CFN at the KIT. Also, we thank Peter
W\"olfle for useful discussions. 
\end{acknowledgments}
\vspace*{-0.5cm}

\end{document}